\documentclass[preprint,showpacs,preprintnumbers,amsmath,amssymb,aps,showkeys]{revtex4}

\usepackage{graphicx}
\usepackage{color}
\usepackage{dcolumn}
\usepackage{bm}
\usepackage{textcomp}

\begin{document}

\title{Acoustic, thermal and flow processes in a water filled nanoporous glasses by time-resolved optical spectroscopy}

\author{R. Cucini$^{1}$, A. Taschin$^{1}$, P. Bartolini$^{1}$ and R. Torre$^{1,2}$}

\affiliation{
$^1$European Lab. for Non-Linear Spectroscopy (LENS), Univ. di Firenze, {\em Via N. Carrara 1, I-50019 Sesto Fiorentino, Firenze, Italy.}
\\$^2$Dip. di Fisica, Univ. di Firenze,{\em Via Sansone 1, I-50019 Sesto Fiorentino, Firenze, Italy.}}

\date{\today}

\begin{abstract}
We present heterodyne detected transient grating measurements on water filled Vycor 7930
in the range of temperature $20-90~^\circ C$. This experimental investigation enables to
measure the acoustic propagation, the average density variation due the liquid flow and
the thermal diffusion in this water filled nano-porous material. The data have been
analyzed with the model of Pecker and Deresiewicz which is an extension of Biot model to
account for the thermal effects. In the whole temperature range the data are
qualitatively described by this hydrodynamic model that enables a meaningful insight of
the different dynamic phenomena. The data analysis proves that the signal in the
intermediate and long time-scale can be mainly addressed to the water dynamics inside the
pores. We proved the existence of a peculiar interplay between the mass and the heat
transport that produces a flow and back-flow process inside the nano-pores. During this
process the solid and liquid dynamics have opposite phase as predicted by the Biot theory
for the slow diffusive wave. Nevertheless, our experimental results confirm that
transport of elastic energy (i.e. acoustic propagation), heat (i.e. thermal diffusion)
and mass (i.e. liquid flow) in a liquid filled porous glass can be described according to
hydrodynamic laws in spite of nanometric dimension of the pores. The data fitting, based
on the hydrodynamic model, enables the extraction of several parameters of the
water-Vycor system, even if some discrepancies appear when they are compared with values
reported in the literature.

\end{abstract}

\pacs{78.47.jj, 47.61.-k, 62.80.+f, 47.56.+r} 

\maketitle

\section{Introduction}

 The study of transport phenomena in heterogenous media is a
fundamental issue of the material science~\cite{Torquato2006,Sheng2006}, its relevance
spans from the basic physics (i.e. the proper definition of the equations of motion
describing the transport processes) to the more recent technological applications.
During the recent years, these studies have to face the scaling down of heterogeneity
towards the nano-metric dimension.

Between the numerous random heterogenous media the two phase solid-liquid materials
represent one of the most investigated. A large impulse, to these studies, has been
given by the petroleum industry aimed at understanding the transport phenomena in
sediment sands and porous rocks. In the basic research, probably the most studied
materials are the solid porous matrix filled by a molecular liquid (e.g. liquid filled
porous glasses), since typically they are well parameterized systems.

A part from the electric processes, three main transport phenomena are generally to be
considered in the solid-liquid heterogenous materials: propagation of acoustic waves,
thermal diffusion and viscous flow (i.e. transport of elastic energy, heat and mass).
These phenomena are generally characterized by very different time/frequency scales so
that often they have been considered as quasi-independent processes, both from the
experimental and theoretical point of view. Moreover, they can be surprisingly described
by relatively simple hydrodynamic models.

The propagation of sound in the liquid-filled porous materials has been described by
phenomenological models based on hydrodynamic/elastic equations for the liquid/solid
phases. The first model has been introduced by M. A. Biot in an important theoretical
work in 1956~\cite{Biot1956a,Biot1956b}. An extraordinary result of the Biot theory is
the prediction of a new slow longitudinal wave besides the usual longitudinal and
transverse waves. This wave has a velocity lower than the one of the bulk
liquid~\cite{Plona1980,Smeulders2005}. In a recent experimental work the validity of this
model in the high frequency range has been investigated~\cite{Taschin2008,Cucini2007},
measuring the hypersonic sound propagation in a nano-porous media.

The viscous flow of liquids through porous media has always been a subject of intense
study~\cite{Bear1972}. Many studies have been concerned with the check of the validity
of the macroscopic law of Darcy. Although Darcy's law was determined phenomenologically,
it is a direct result of the Navier-Stokes hydrodynamic equations~\cite{Neuman1977},
thus checking its validity is a straightforward analysis of the hydrodynamical character
of liquid flowing. Darcy's law is shown to be valid for a wide range of porous media
from porous rocks, sands~\cite{Bear1972} and glass beads \cite{Chandler1981} whose
heterogeneities are micrometric, up to porous glasses like Vycor which have
heterogeneities of the order of few nanometers
\cite{Debye1959,Lin1992,Li2000,Vadakan2000,Vadakan2000b}. Recently, many experimental
and simulation works aimed at understanding to what extent the Navier-Stokes equation
can describe the liquid flow at decreasing of the confinement
size~\cite{Travis1996,Spohr1999,Li2000,Huang2007}.

The thermal conductivity of heterogenous media is a complex problem which has to be
characterized in a general theoretical framework~\cite{Torquato2006}. The prediction of
the effective permeability of saturated porous media remains, despite of many
experimental and theoretical works, an unsolved problem in heat transfer science. In
particular this quantity depends, as well as on macroscopic parameters of the two
constituent phases, also on the particular morphology of the porous material which is in
general experimentally not accessible. Nevertheless, the theory is able to fix the limits
of the effective conductivity and give several approaches for calculating this quantity
in particular kind of porous media~\cite{AdHT2006}.

Though the transport phenomena in liquid-filled porous glasses has been previously
studied in the literature, many basic questions remain open. In our opinion, one of the
more relevant is at which extent the hydrodynamic models are valid, especially when the
media heterogeneity scale down versus molecular length scales.

We think that other experimental researches could infer new information, especially with
techniques poorly applied in this field, like the time resolved spectroscopy
experiments. Transient grating (TG) experiments~\cite{Eichler1986,Yan1995} are powerful
tools for investigate the relaxation dynamics of complex
liquids~\cite{Yan1995,Torre2001,Taschin2006,CapBook_TG2008}, but only few previous
experimental works utilized these techniques to investigate the porous glass
samples~\cite{Ritter1988,Taschin2007,Cucini2007,Taschin2008}. A particular kind of TG
experiment offers the possibility to study, at one time, the acoustic propagation, the
liquid flow and the thermal diffusion processes. The very broad time window covered by
this experiment, typically from $10^{-9}$ to $10^{-3}~s$, gives access to a dynamic
range hardly explored by other methods.

In this paper we present heterodyne detected transient grating (HD-TG) measurements on
water filled Vycor nano-porous glass. The measured signal shows aspects related to the
acoustic waves propagation, to the viscous liquid flow through pores, and, lastly, to
the thermal diffusion. The transient grating experiment reported here is shown to be a
powerful tool to measure the transport processes in liquid filled porous glasses.

To analyze the data we employ an hydrodynamic model introduced by Pecker and Deresiewicz
\cite{Pecker1973} which is a extention of the Biot's theory~\cite{Biot1956a,Biot1956b}
to include the thermal effects. The model enables a safe and meaningful addressing of
the different contributions present in the HD-TG signal, describing them on the base of
few parameters characterizing the heterogenous system.

The present work is divided as follows. In Sec. \ref{sec_TG_exp} we report an
overview of the TG experimental technique describing the involved excitation mechanisms
and the measured dynamics with probing process. In Sec. \ref{sec_TTGE} we present the
experimental aspects of the work, laser systems, experiment optical set-up and the sample
preparation. Section \ref{sec_Exp_Res} is devoted to the presentation of the data
while the subsequent one to the introduction of the theoretical model used to analyze
these data. Finally, in the last section we show the data analysis and we discuss the
obtained results.

\section{Transient grating experiments}\label{sec_TG_exp}
\begin{figure}[t]
    \centering
    \includegraphics[scale=0.9]{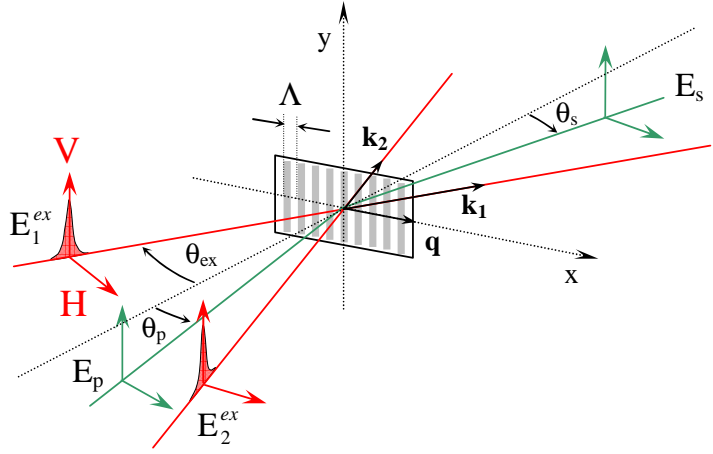}
    \caption{Schematic drawing of a transient grating experiment.
    Two excitation pulses, $E^{ex}_1 $ and $E^{ex}_2$ induce
    an impulsive spatial modulation of the dielectric constant with step $\Lambda$.
    The relaxation of the induced transient modulation is probed by
    the Bragg scattering of a third beam, $E_{p}$.}
    \label{TG_setup}
\end{figure}

In a TG experiment, two infrared laser pulses, obtained dividing a single pulsed laser
beam, interfere within the sample and produce an impulsive spatially periodic variation
of the dielectric constant. The spatial modulation is characterized by a wave vector
$\mathbf{q}$ which is given by the difference of the two pump wave vectors
$\mathbf{k}_{1}-\mathbf{k}_2$ (see Fig.~\ref{TG_setup}). Its modulus is
$q=4\pi\sin(\theta_{ex})/\lambda_{ex}$, where $\lambda_{ex}$ and $\theta_{ex}$ are the
wavelength and the incidence angle of the exciting pumps, respectively. The relaxation
toward equilibrium of the induced modulation can be probed by measuring the Bragg
scattered intensity of a second cw laser beam. The time evolution of the diffracted
signal supplies information about the dynamic of the relaxing TG and, consequently, on
the dynamical properties of the studied sample.

 TG experiments fall within the framework of the four-wave-mixing
theory~\cite{Shen1984,Eichler1986,CapBook_TG2008,Hellwarth1977}. Under a few
approximations, in particular, if the laser pulses do not have any electronic resonance
with the material, it can be proved that the TG experiment can be divided in two
separated processes: excitation and probing. Moreover, when the heterodyne detection is
employed, the signal turns out to be directly proportional to the dielectric constant
change induced by the pumps, $\delta \epsilon_{ij}$, or to the response function of the
system, $R_{ijkl}$~\cite{CapBook_TG2008}:
\begin{equation}
    S^{HD}(q,t)\propto \delta \epsilon_{ij}(q,t)\propto R_{ijkl}(q,t)\label{TG-sign}
\end{equation}
where the cartesian indexes $i$ and $j$ define the polarizations of the diffracted and
the probe fields, $k$ and $l$ define the polarizations of the pump fields. Both $\delta
\epsilon_{ij}$ and $R_{ijkl}$ are spatial Fourier components corresponding to the
exchanged wave vector $\mathbf{q}$, i. e. the grating wave vector. Hence, the heterodyne
detected TG signal measures directly and linearly the relaxation processes defined by the
tensor components of the response function. By selecting different polarizations of the
fields, different elements of the response function tensor are probed. Generally, these
elements can be different~\cite{Taschin2001,Pick2004,Azzimani2007a,Azzimani2007b}. In
materials, where the coupling between translational and rotational degrees of freedom is
strong, the birefringence effects can be relevant and the TG signal will depend on the
field polarization. When, instead, all the birefringence contributions are negligible,
the signal turn out to be independent on the field polarization and the induced
dielectric constant tensor is simple $I\delta\epsilon$, where $I$ is the identity
tensor~\cite{CapBook_TG2008}.

Without birefringence effects and in bulk homogeneous materials, the transient grating is
mainly induced by two effects. The electrostriction and the heating. The first effect
arises from the interaction between the molecular dipoles induced by the pump fields and
the same pump fields. The electrostriction produces a pressure grating and consequently a
density grating launching two counter-propagating acoustic waves whose superposition
makes a standing acoustic wave. The second exciting process arises from a weak absorption
of the pump infrared radiation which is resonant with some vibrational states. The
increasing of energy due to the pump absorption, generally, quickly thermalizes by fast
non radiative channels and builds up a temperature grating. This one produces a pressure
grating and then the pressure grating produces a density grating via thermal expansion.
Now, besides acoustic waves, we have more a constant density grating supported by the
temperature grating which relaxes by thermal diffusion. As we shall see in section
\ref{sec_Exp_Res}, in a heterogeneous material the excitation sources are the same
but the effects can be more complex.

\section{Experimental procedures}\label{sec_TTGE}
\begin{figure}[t]
   \centering
        \includegraphics[scale=0.6]{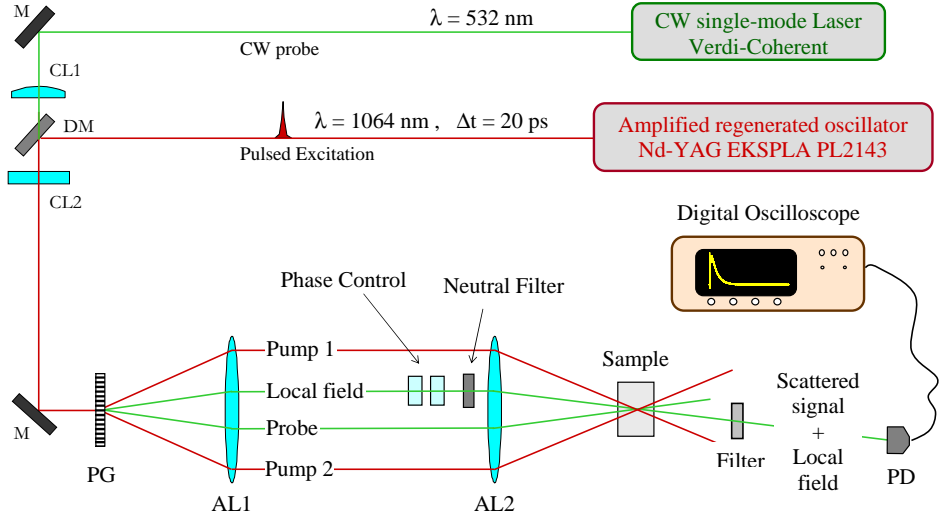}
        \caption{Optical set-up and laser system for the TG experiment with optical
        heterodyne detection. M are the mirrors; CL\# the
        cylindrical lenses; DM a dichroic mirror; PG is the phase grating; AL\# the
        achromatic lenses and PD is the photodiode.}
        \label{setup}
    \end{figure}

A detailed description of the lasers and the optical set-up of our TG experiment is
reported in references~\cite{CapBook_TG2008}. Here we just recall the main aspects. A
sketch of the TG optical set-up is reported in Fig.~\ref{setup}. The infrared pump
pulses have a $1064~nm$ wavelength, a temporal length of $20~ps$ and repetition rate of
$10~Hz$. They are produced by an amplified regenerated oscillator (Nd-YAG EKSPLA PL2143).
The typically used pump pulse energy was $5~mJ$. The probing beam, instead, is a
continuous-wave laser at $532~nm$ produced by a diode-pumped intracavity-doubled Nd-YVO
(Verdi-Coherent). The two laser beams are collinearly sent to a phase grating (PG) to get
the two pump pulses and the probe and local field beams. These are obtained taking the
$+1$ and $-1$ diffraction orders of the infrared and green lasers. A couple of confocal
achromatic lenses, AL1 and AL2, collects all the four beams and focuses them on the
sample. The phase grating directly supplies a probe at the right Bragg angle and a local
field exactly collinear with the scattered field and phase locked with the probe. The
HD-TG signal is optically filtered and measured by a fast avalanche silicon photodiode
with a bandwidth of 1 GHz (APD, Hamamatsu). The signal is then amplified by a
DC-$800~MHz$ AVTECH amplifier and recorded by a digital oscilloscope with a $7~GHz$
bandwidth and a $20~Gs/s$ sampling rate (Tektronix). The instrumental function of our
setup has a temporal full width half maximum of 1 $ns$ and is mainly determined by the
bandwidth of the detector and its amplifier~\cite{CapBook_TG2008}.

Vycor 7930 is an open cell porous glass which is produced by a spinodal demixing in a
borate-rich and borate-poor glass and subsequent bleaching of the borate-rich phase~\cite{Corning2008}.
It has nominal values of porosity and mean pore size of
$28\%$ and $4~nm$ respectively. The solid constituent of Vycor 7930 is the glass Vycor
7913 which is composed by $96\%$ of silica, by $3\%$ of boron oxide and the remaining
part mostly by aluminum oxide and zirconium oxide. Our sample has been purchased from
``Advanced Glass and Ceramics'' in a cylindrical form, with a diameter of $15~mm$ and
$10~mm$ thickness. In order to remove all the absorbed organic elements we washed the
sample in $30\%$ water-hydrogen peroxide solution and then we heated it in a muffle up to
$800~^\circ C$ at a rate of $0.3~K/min$. The sample has been kept at this temperature for
several hours. Water has been obtained from a double-distilled water vial prepared for
pharmaceutical purposes. The filled matrices were, then, placed in a thin teflon tube and
closed between two circular quartz windows. The whole was inserted in a cylindrical
copper cell (similar to the one reported in~\cite{Halalay1990}). The holder was then
stabilized in temperature with a stability of $\pm0.1~K$.

\section{Experimental Results}\label{sec_Exp_Res}
\begin{figure}[t]
\centering
    \includegraphics[scale=0.55]{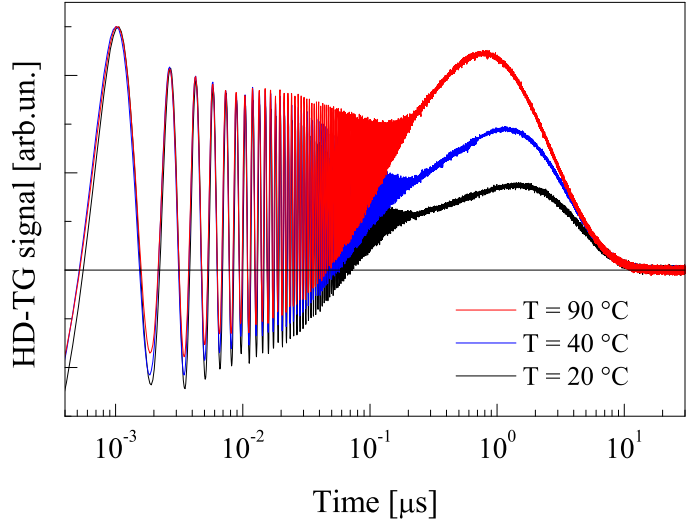}
    \caption{HD-TG data on Vycor-water system at $q=1~\mu m^{-1}$ at three different temperatures.
    At short times, the data show damped acoustic oscillations,
    at middle times, a density rearrangement related to liquid flow and finally,
    at long times, the data show a decay due to the thermal diffusion.
    The acoustic waves are practically induced only by the electrostrictive effect,
    while the remaining part of the signal by the thermal excitation.}
    \label{data}
\end{figure}

The HD-TG signals on water-filled Vycor have been collect in the range of temperature
$20-90~^\circ C$ at different wave vectors. The data did not show a dependence on the
polarizations of beams and consequently birefringence contributions were
negligible~\cite{Taschin2006}. Hence, we recorded the data in a single configuration of
beams polarization in which all the beams had a polarization perpendicular to the
scattering plane as sketched in Fig.~\ref{TG_setup}. The data at three different
temperatures, 90, 40 and 20 $^\circ C$, are shown in Fig.~\ref{data} (following the
usual TG signal convention, the sign of the data is chosen positive for a negative change
of density~\cite{CapBook_TG2008}).

Before explaining the data features, some considerations have to be discussed. As just
mentioned in section~\ref{sec_TG_exp}, in a TG experiment we have two exciting
sources: the electrostrictive effect and the heating process. In the water filled Vycor,
the first excitation is surely present in both materials, even if the electrostrictive
strength could be different in the two materials, definitely we expect that this
excitation source launches the acoustic waves. Differently, the heating should be much
weaker in the solid part than in the liquid water or at the pore surfaces. In fact, the
liquid water has a no negligible absorption coefficient at the pump wavelength and also
the pore surfaces present a weak infrared absorption due to the presence of the silanol
and boranol groups (Si-OH and B-OH). Nevertheless, on a very fast time scale, in the
illuminated areas, all the sample components (solid part of the matrix, pore surfaces and
liquid water) reach the same temperature. This makes applicable the assumption of
\emph{local thermal equilibrium}, i.e. we are supposing that the liquid and matrix
temperatures are always locally equilibrated. Kaviany (\cite{Kaviany1995} pp. 120-121)
reports criteria of validity for the local thermal equilibrium approximation which are
well fulfilled for the water-Vycor system. Namely, the approximation of local thermal
equilibrium requires that the time scale associated with the inter-pore heat transfer
must be much smaller than the time scale associated with the macroscopic heat transfer
experimentally measured. The heat transfer between two pores occurs on a nanometric
length scale while the macroscopic length scale is given by the spacing of the induced
grating, $\Lambda=2\pi/q=6.28~\mu m$, for $q=1~\mu m^{-1}$. Moreover, the interphase heat
transfer is a very efficient process thanks the very large specific pore surface area. We
expect this process to evolve on shorter time scales than the time scales measured in the
HD-TG signal. So, also in our heterogenous sample the pump heating produces an uniform
temperature grating as in a bulk homogeneous sample.

Now, other important aspects have to be considered to understand the HD-TG signals: 1)
the thermal expansion coefficient of the matrix is very low compared to the water one; 2)
the bulk modulus of the matrix is higher than the water one, the matrix is stiff. Since
the thermal expansion of water would be greater than that of the matrix, at first the
expansion of the liquid is reduced by the stiffness of the matrix which exerts a pressure
on the liquid. At later times, the liquid flows through the pores to nullify the matrix
pressure which is at first positive and then negative, as it will be proved later using a
data simulation based on the hydrodynamic model. So, the liquid can practically expand
only via the flowing. This implies also that the acoustic oscillations induced by the
temperature grating are characterized by a very low amplitude.
\begin{figure}[t]
\centering
    \includegraphics[scale=0.53]{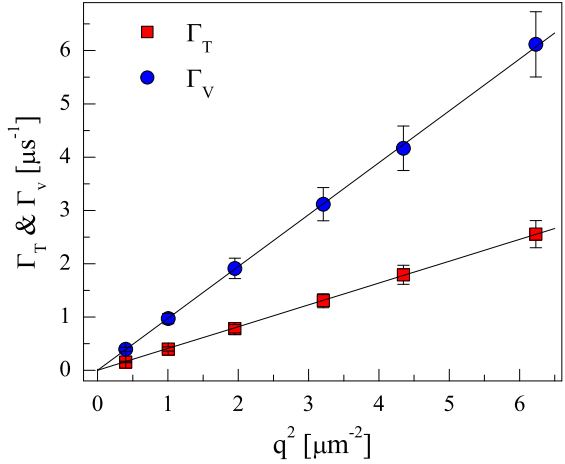}
    \caption{Decay constants of the two exponentials describing the long part of the signal as a
    function of $q^2$ at the fixed temperature of 30 $^\circ C$.
    $\Gamma_V=1/\tau_V$ and $\Gamma_T=1/\tau_T$ are, respectively, the decay constants of the exponential rise and fall of the signal.
    The linear behavior confirms the diffusive character of the viscous
    flow mode and of the thermal diffusion.}
    \label{tau_vq2}
\end{figure}

For the aforementioned reasons, the data show, at short times, damped acoustic
oscillations induced almost only by the electrostrictive effect, at intermediate times, a
density rearrangement related to liquid flow inside the pores.

From preliminary fits of the intermediate and long parts of the signals, we obtain that
the rise is well reproduced by a single exponential whose time constant, $\tau_V$, shows
a $q^{-2}$ dependence. It is, therefore, a diffusive mode and it turns to be the Biot
slow wave in diffusive regime, as it will shown later using the hydrodynamic model. The
time constant, then, is related to the constant diffusion of the Biot slow wave
(hydraulic diffusivity coefficient) which is connected to Vycor permeability and liquid
viscosity. Clearly, in this case the temperature grating plays an important role, it is
the pumping source for the liquid motion. In Fig.~\ref{tau_vq2} we report the inverse
of constant times of the exponential signal rise, $\Gamma_V=1/\tau_V$, and the
exponential signal fall, $\Gamma_T=1/\tau_T$, as a function of the $q^2$ for the
temperature fixed at 30 $^\circ C$. The probed wave vectors were $q=0.63$, 1.00, 1.39,
1.76, 2.15, and 2.51 $\mu m^{-1}$. The linear behavior of the time constants confirms the
diffusive character of the modes.

\section{Theoretical background}\label{sec_ThBkg}
An explicit function of the TG signal can be obtained developing the dielectric function
$\delta\epsilon$ through the hydrodynamic variables which, in the absence of
birefringence effects, are only the density and the temperature. According to a
first-order approximation, the dielectric constant change in a homogeneous bulk material
is~\cite{Pecora1976}:
\begin{equation}
    \delta \epsilon(t)=\left( \frac{ \partial \epsilon }
    {\partial \rho }\right)_{T}\delta \rho(t) +
    \left( \frac{\partial \epsilon }{\partial T}\right)_{\rho }\delta
    T(t),\label{DroDT1}
\end{equation}

Now we have to consider that our sample is composed by two interconnected materials,
Vycor and water. The dielectric constant change will depend on two densities and two
temperatures:
\begin{eqnarray}
    \delta \epsilon(t) &= \left( \frac{\partial \epsilon}{\partial \rho_1} \right) _{T_m,T_l,\rho_2}\delta
    \rho_1(t)
    +\left( \frac{\partial \epsilon }{\partial \rho_2}\right)_{T_m,T_l,\rho_1}\delta\rho_2(t) \nonumber\\
    &+ \left( \frac{\partial \epsilon}{\partial T_m} \right) _{T_l,\rho_1,\rho_2}\delta
    T_m(t)
    +\left( \frac{\partial \epsilon }{\partial T_l}\right)_{T_m,\rho_1,\rho_2}\delta
    T_l(t)
    \label{DroDT2}
\end{eqnarray}
where $\rho_1=(1-\phi)\rho_s$ and $\rho_2=\phi\rho_l$ are the average densities of matrix
and water respectively, in which $\phi$ is the porosity of Vycor, $\rho_s$ the density of
solid and $\rho_l$ the water density. $T_m$ and $T_l$ are the matrix and water
temperatures respectively. The dielectric constant change induced directly by the
temperature variation, in Eq.~(\ref{DroDT1}), is generally neglected in the TG signal
analysis because for most of liquids $(\partial \epsilon /\partial T)_{\rho}\delta T$ is
much lower than $(\partial \epsilon /\partial \rho)_{T }\delta \rho$. Water is an example
for which this condition is not satisfied, and the TG signal depends also on this
contribution~\cite{Taschin2006}. Thus, we can neglect the matrix temperature
contribution, but not the water temperature contribution that could have an important
weight:
\begin{equation}
    \delta \epsilon(t) = \left( \frac{\partial \epsilon}{\partial \rho_1} \right)\delta
    \rho_1(t)
    +\left( \frac{\partial \epsilon}{\partial \rho_2}\right)\delta\rho_2(t)
    +\left( \frac{\partial \epsilon}{\partial T_l} \right)\delta T_l(t)
    \label{DroDT3}
\end{equation}
The previous equation can be rewritten in the following way:
\begin{equation}
    \delta \epsilon(t)=\left( \frac{\partial \epsilon}{\partial \rho_2}\right)
    \left[A_{12} \delta \rho_1(t)+\delta\rho_2(t)+E\frac{\rho_2}{T_0}\delta T_l(t)\right]
    \label{DroDT4}
\end{equation}
where we have defined
\begin{eqnarray}
    A_{12}=\frac{\left( \partial \epsilon /\partial \rho_1\right)}{\left(\partial
    \epsilon /\partial\rho_2\right)}\quad  and \quad
    E=\frac{T_0 \left(\partial \epsilon/\partial T_l\right)}{\rho_2\left(\partial \epsilon / \partial \rho_2 \right)}
\end{eqnarray}
$A_{12}$ and $E$ are amplitude parameters which define the relative weight between the
three different contributions in the final signal. $E$ is defined similarly
in~\cite{Taschin2006}, but its value could be different from that measured for the bulk
water because of the strong interactions between water molecules and the hydrophilic
surfaces of Vycor pores. This interaction strongly modifies the hydrogen bonds among
water molecules and it could change the value of $E$. Thus, both the parameters, $A_{12}$
and $E$ must be valued from the data fitting.

\subsection{Hydrodynamic model}

The time evolution of the two densities and the two temperatures can be obtained from
the thermo-poroelastic model introduced in 1973 by Pecker and Deresiewicz (P-D)
~\cite{Pecker1973}. This is an extension of the well known Biot model on the wave
propagation on poroelastic system to account for the temperature effects. An earlier
attempt to include the temperature was made by Zolotarev in 1965~\cite{Zolotarev1965},
but here strong approximations were introduced. The model of Pecker and Deresiewicz has
been later deeply revisited as linearized approximation of more complex models by
Gajo~\cite{Gajo2002} and Youssef~\cite{Youssef2007}. We shall follow here the Pecker and
Deresiewicz (P-D) model adopting the same notation since exactly alike to the Biot one.

The main assumptions of the theory are the same of the Biot theory. In particular: the
solid phase of the system is perfectly elastic, homogeneous and isotropic; the liquid is
a compressible perfect fluid; the liquid viscosity is only introduced into a dissipation
function to account for the sound wave attenuations due to the viscous friction between
matrix and liquid; sound wavelengths are much larger than the heterogeneity dimensions;
finally, all the transformations (displacements, strains, etc.) are considered
infinitesimal in order to obtain a linear theory.

In comparison with Biot model, two relaxation equations are added for the matrix and
liquid temperatures and some thermo-mechanical coupling terms are inserted in all the
equations to account for the coupling effects between the two temperatures and the two
densities.

Since in an isotropic medium TG experiment is sensible only to density and temperature
changes, we focus our attention only to the motion equations of the matrix and liquid
dilatations discarding the shear motions.

Defining with $\mathbf{u}$ and $\mathbf{U}$ the displacements of solid and liquid phases
\cite{Landau1959}, the dilatations are defined by $e=\mathbf{\nabla}\cdot\mathbf{u}$ and
$\mathcal{E}=\mathbf{\nabla}\cdot\mathbf{U}$. $e$ and $\mathcal{E}$ are related to the
matrix and liquid density changes, $\delta\rho_1$ and $\delta\rho_2$, by the following
identities:
\begin{equation}
    \delta\rho_1=-\rho_1 e \quad and \quad \delta\rho_2=-\rho_2 \mathcal{E}\label{eE_den}
\end{equation}
The time evolution of the dilatations $e$ and $\mathcal{E}$ and the temperature
variations $\delta T_m$ and $\delta T_l$ can be evaluated by the P-D model. In the
appendix we report the detailed description of the P-D equations and all the expression
of the coefficients. Few approximations suitable for the water-Vycor system can be made
to this model (see appendix), in particular we can assume the \emph{local thermal
equilibrium} by which the liquid and matrix temperatures can be always considered locally
equilibrated: i.e. $T_m=T_l=T$. After that, the linearized P-D equations in the Fourier
q-space become:
\begin{eqnarray}
    \rho_{11}\ddot{e}+\rho_{12}\ddot{\mathcal{E}}
    +b(\dot{e}-\dot{\mathcal{E}})+q^2Pe+q^2Q\mathcal{E}-q^2R_1\delta T=0 \nonumber\\
    \rho_{12}\ddot{e}+\rho_{22}\ddot{\mathcal{E}}-b(\dot{e}-\dot{\mathcal{E}})+q^2Qe+q^2R\mathcal{E}-q^2R_2\delta T=0 \nonumber\\
    F\dot{\delta T}+R_1T_0\dot{e}+R_2T_0\dot{\mathcal{E}}+q^2k \delta T=0 \label{Pecker0}
\end{eqnarray}
where the densities $\rho_{ij}$ are related to the solid and liquid densities, $\rho_s$
and $\rho_l$, by $\rho_{11}=(1-\phi)\rho_s-\rho_{12}$, $\rho_{22}=\phi\rho_l-\rho_{12}$
and $\rho_{12}=(1-\tau)\phi\rho_l$ where $\tau$ is the tortuosity of the
matrix~\cite{Biot1956a,Biot1956b}. The $b(\dot{e}-\dot{\mathcal{E}})$ is a viscous
friction term arising from the relative motion between the two materials, being
$b=\phi^2\mu_l/k_D$, with $\mu_l$ the water dynamic viscosity and $k_D$ the Vycor
permeability. This term is, at the same time, the damping source for the acoustic waves
and the hindrance to the liquid slow flow through pores. $P$, $Q$ and $R$ are the
generalized isothermal elastic moduli, the coefficients $R_{i}$ are the generalized
expansion coefficient and the $F$ coefficient is a generalized specific
heat~\cite{Biot1956a,Biot1956b,Johnson1982,Carcione2001}. $k$ is the effective thermal
conductivity. The complete definition of these coefficients is reported in appendix.
These equations enable to calculate the $q$-component of the time dependent variables
$e$, $\mathcal{E}$ and $\delta T$, and thus, using the equations ~\ref{TG-sign},
~\ref{DroDT4} and~\ref{eE_den}, to simulate the measured signal.

The model of P-D model is an extension of Biot's model in the low frequency limit, i.e.
is valid for motions at frequencies lower than the Biot characteristic frequency. This
quantity is defined by $f_{c}= \mu_{l} / \pi\rho_{l}r^{2}$ with $r$ the mean pore radius
of the matrix. For the water-Vycor system the characteristic frequency is $26-80~GHz$ in
the temperature range considered. This value has to be compared to the higher frequency
experimentally excited which is almost 0.6 $GHz$ (the experimental frequency can be
obtained by $f=cq/2\pi$ where $c=4~Km/s$ is the measured sound velocity and $q=1~\mu
m^{-1}$ the exchanged wave vector). Thus, our experiment is testing dynamics at
frequencies always lower then $f_{c}$ and the present model is theoretically suitable to
describe our signals. Moreover, when the involved frequency is much lower than the
characteristic frequency, the dependence on the tortuosity, $\tau$, of the solutions of
Eq.s (\ref{Pecker33}), should be negligible. This condition has been verified in our case
by the fitting procedure. As final consideration, we want to stress that the above
equations, and consequently their solutions, do not depend explicitly on the pore
dimension. Nevertheless, this parameter enters in the definition of the characteristic
frequency, thus, fixing the applicability limit of the theory.

\subsection{TG response and signal}

The set of Eq.s~(\ref{Pecker0}) can be reduced to a first order differential equation
system by introducing two additional variables $\psi_1=\dot{e}$ and
$\psi_2=\dot{\mathcal{E}}$. The set of equations (\ref{Pecker0}) can, thus, be easily
written in the compact form
\begin{equation}
    \dot{\mathbf{X}}(t)=-\mathbf{M}\cdot\mathbf{X}(t)\label{eq_matrix}
\end{equation}
where $\mathbf{X}=(e,\mathcal{E},\psi_1,\psi_2,\delta T)$ and M is a matrix of
coefficients which can be easily obtained from Eq.s~(\ref{Pecker0}). Indeed we can
solve the system by
diagonalizing, with standard routines, the non symmetric $\mathbf{M}$ matrix, i.e. $\mathbf{M}%
\cdot\mathbf{V=V}\cdot\mathbf{D}$ where $\mathbf{D}$ is diagonal and the resulting time
solution is written as
\begin{equation}
    \mathbf{X(}t\mathbf{)=V}\cdot\exp(-\mathbf{D\ }t\mathbf{)\cdot}\left[
    \mathbf{V}^{-1}\cdot\mathbf{X}(0)\right] \label{TG_eqsol}%
\end{equation}
We note that it is possible to avoid the $\mathbf{V}$ matrix inversion, since
$\mathbf{Y=V}^{-1}\cdot\mathbf{X(}0\mathbf{)}$ is the solution of linear system
$\mathbf{V}\cdot\mathbf{Y=X(}0\mathbf{)}$. We want to stress that only the elements of
the amplitude matrix, $\mathbf{V}$, and the root matrix, $\mathbf{D}$, need to be
numerically calculated. Each element of the solution vector $\mathbf{X}(t)$ is a sum of
two oscillating terms, cosine and sine rising from the two complex and conjugate roots of
$\mathbf{D}$, and three pure exponential terms deriving from the other three real roots
of $\mathbf{D}$.

The temporal expressions of the dilatations and the temperature must be calculated
considering that the electrostriction produces a non zero initial condition on $\psi_1$
and $\psi_2$ and the heating produces a no null initial condition on $\delta T$:
$\mathbf{X}(0)=(0,0,\psi_1(0),\psi_2(0),\delta T(0))$. These solutions together with the
expressions for the densities (\ref{eE_den}) and the expression for dielectric constant
change (\ref{DroDT4}) give the signal function:
\begin{eqnarray}
\nonumber S^{HD}(t) \propto \delta \epsilon(t)=&Ae^{-t/\tau_S}cos(\omega_S t) +
Be^{-t/\tau_S}sin(\omega_S t)+ \nonumber\\
&Ce^{-t/\tau_1}+De^{-t/\tau_2}+Fe^{-t/\tau_3}\label{signal-final}
\end{eqnarray}
where the amplitudes $A$, $B$, $C$, $D$, and $F$, the acoustic parameters $\tau_S$
and $\omega_S$, and the time constants $\tau_1$, $\tau_2$, and $\tau_3$ are functions
depending on all the hydrodynamic parameters appearing in the starting equations and
on the initial conditions and need to be numerically calculated at changing of the
fitting parameters listed below.

The oscillating terms in Eq.~(\ref{signal-final}) accounts for the induced acoustic
oscillations and the three exponentials for the remaining part of the signal. As
previously discussed, the acoustic oscillations induced by the temperature grating are
partially prevented. Anyhow, these ones have been equally considered in the fitting
analysis and they proved to be very small at all the temperatures, around 5$\%$ of those
induced by the electrostriction.

Finally, Eq. (\ref{signal-final}) with the addition of electronic response function
and convoluted with the experiment instrumental function, yields the used fitting
function~\cite{CapBook_TG2008}.

\section{Data analysis and discussion}\label{sec_Data_Anal}

The literature provides many of the water and Vycor data appearing in the P-D equations:
the densities $\rho_s$, $\rho_l$, the expansivities $\alpha_s$, $\alpha_l$, the water
viscosity $\mu_l$, the specific heats $C_{ps}$, $C_{pl}$, the Vycor porosity $\phi$, the
transverse and longitudinal sound velocities of solid constituent for the calculus of
$K_s$ and the adiabatic sound velocity of water and the specific heat ratio $\gamma_r$
for the calculus of $K_l$.

The other parameters were free fitting parameters. In table~\ref{tb_par} we list the
parameters locked to the literature values with the corresponding references.

As reported in Sec. \ref{sec_TTGE}, the solid constituent of Vycor 7930 is not pure
silica but Vycor 7913, a mixture of $96\%$ of silica and $3\%$ of boron oxide. Corning
company~\cite{Corning2008} supplies us all the parameters of interest of Vycor 7913 only
at room temperature. To overcome this problem, we made the hypothesis that the
thermodynamic parameters of Vycor 7913 follow the temperature dependence as in
silica~\cite{Scopigno00}. Considering the restricted range of temperature analyzed, this
is surely a valid approximation.

For the calculation of the shear and bulk moduli of dry matrix, we take the matrix
longitudinal sound velocity as free fitting parameter and the transverse sound velocity
proportional to the longitudinal one as in the solid phase: $V_{t,m}=0.62 V_{l,m}$ where
the $m$ and $s$ indices refer to the matrix and solid materials and $l$ and $t$ to the
longitudinal and transverse wave mode
respectively~\cite{Terki1998,Caponi2002,Levelut2007}.

As porosity, we use the value of $31\%$, which is the mean of values reported in the
literature~\cite{Debye1959,Vadakan2000,Vadakan2000b,Gille2002}.

In reference~\cite{Taschin2008} we proved that the Biot theory is not able to predict the
acoustic damping times of TG data giving far overestimated values. The reason, probably,
lies in the only damping mechanism considered in the theory, i.e. the energy dissipation
due to the viscous friction between the matrix and liquid. Other damping effects, like
the intrinsic dissipation of two materials are not included. P-D model does not introduce
substantial modifications for the wave propagation and retains the same limitations of
the Biot model. In order to avoid that the extraction of other interesting parameters
could be affected by this mismatch, we multiply the oscillating terms of the solution
(\ref{signal-final}) by $e^{-t/\tau_A}$ to account for the real exponential decay of the
acoustic waves.

Finally, the free fitting parameters were: the three initial conditions $\psi_1(0)$,
$\psi_2(0)$ and $\delta T(0)$, the relative amplitudes $E$ and $A_{12}$, the longitudinal
sound velocity of the matrix $V_{l,m}$, the acoustic damping time $\tau_A$, the
permeability $k_D$ and, lastly, the effective thermal conductivity $k$. This last
parameter could to be fixed to the value given by the expression used in the the P-D
model $k=(1-\phi)k_s+\phi k_l$, where $k_s$ and $k_l$ are the solid and liquid thermal
conductivity, respectively. According to our data fitting, this value turned out not to
be correct and we were forced to leave it free.

\begin{table*}[tbp]
\caption{Water and Vycor parameters fixed in the fitting procedure to the values found in
the corresponding references.} \label{tb_par}
\begin{center}
\begin{tabular}{c|l|c|c}
\hline\hline Symbol & Definition & Values @ 20$~^\circ C$ & Ref. \\
\hline
$\rho_s$  & ~solid density & $2180$ $Kgm^{-3}$& \cite{Corning2008}\\
$\rho_l$  & ~water density & 998 $Kgm^{-3}$ & \cite{Handbook70}\\
$\alpha_s$& ~linear solid expansivity & $7.5 \times 10^{-7}$ $K^{-1}$ & \cite{Corning2008}\\
$\alpha_l$&~volume water expansivity& $2.07 \times 10^{-4}$ $K^{-1}$& \cite{Handbook70}\\
$\mu_l$  &~water viscosity & $10^{-3}$ $Pa\cdot s$& \cite{Handbook70}\\
$C_{ps}$  &~solid specific heat & $739$ $J K^{-1}Kg^{-1}$& \cite{Lord1956}\\
$C_{pl}$  &~water specific heat & $4184$ $J K^{-1}Kg^{-1}$& \cite{Handbook70}\\
$\phi$  &~porosity & $31\%$& \cite{Debye1959,Vadakan2000,Vadakan2000b,Gille2002}\\
$V_{l,s}$&~solid longitudinal sound velocity & 5779 $ms^{-1}$ & \cite{Corning2008,Scopigno00}\\
$V_{t,s}$&~solid transverse sound velocity & 3580 $ms^{-1}$ & \cite{Corning2008,Scopigno00}\\
$V_{l}$&~water adiabatic sound velocity& 1483 $ms^{-1}$& \cite{Nist2008}\\
$\gamma_r$ &~water specific heat ratio & 1.0065 & \cite{Nist2008,Johari1996}\\
\hline
\end{tabular}
\end{center}
\end{table*}

\begin{figure}[t]
\centering
    \includegraphics[scale=0.55]{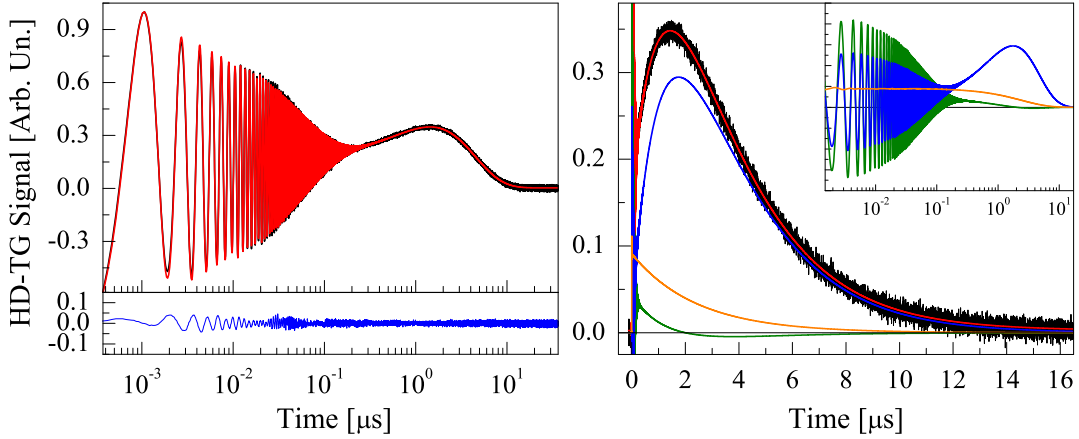}
    \caption{Comparison between data (black line) and fit (red line)
    for the temperature $20~^\circ C$ with the relative discrepancy (left panel).
    In the right panel the three different contributions whose sum gives the fit of the datum at $20~^\circ C$ are shown. These are the matrix density contribution (green line),
    the water density (blue line) and the temperature (orange line) contributions.
    The inset shows the same contributions in x-log scale.}
\label{datafit}
\end{figure}

The used fitting function is able to reproduce the data in all temperature range
analyzed. In the left panel of Fig.~\ref{datafit} we report in semi-log scale the
fit-data comparison of the signal at 20 $^\circ C$ together with the corresponding
discrepancy. In the right panel of the same figure, we show in linear scale the same
fit-data comparison, together with the three different contributions appearing in
expression (\ref{DroDT4}). These are the matrix density contribution (green line), the
water density (blue line) and the temperature one (orange line) contributions. The sum of
these three contributions gives the shown fitting curve. As clearly visible, the water
density change is the main term and it is the only one yielding the bump in the signal.
Thus, the long time dynamics is mainly related to the water density changes and this is
clearly due to the great difference between the water and Vycor expansivities and to the
stiffness of the matrix. The signal contribution induced directly by the temperature
(third term of Eq.~\ref{DroDT3}, $\left({\partial \epsilon}/{\partial T_l}\right)\delta
T_l(t)$) is practically described by a single exponential decay, it does not show any
acoustic oscillation and has an intensity weakly dependent on the temperature. We want to
stress that, as regards the acoustic oscillations, the liquid and matrix move in phase
(see the inset of Fig.~\ref{datafit}). Contrary, in the major part of the signal the
water and matrix densities have opposite derivatives, i.e. when the liquid tends to
expand the matrix tends to compress and vice versa. This implies an out-phase motion of
the liquid and matrix as expected for the slow sound predicted by the Biot theory.

Once we extracted all the parameters for a given temperature, we can use them to
simulate the time dependence of an isolate physical observable. It is interesting to
study the temporal behavior of the liquid pressure inside the pores. This can be
obtained from the constitutive equation for the pore fluid~\cite{Pecker1973}:
\begin{equation}
    \delta p(t)=-\frac{1}{\phi}[Qe(t)+R\mathcal{E}(t)-(R_1+R_2)\delta T(t)]\label{pressure}
\end{equation}
\begin{figure}[t]
\centering
    \includegraphics[scale=0.4]{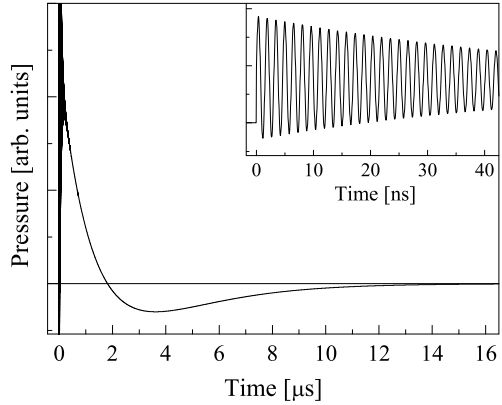}
    \caption{Temporal evolution of the pore pressure obtained from the constitutive
    equations for the pore fluid (Eq. \ref{pressure}) using the parameters extracted
    by the fit of the datum at 20 $^\circ C$. The pressure shows two relaxation modes,
    a fast one which drives the acoustic oscillations (inset in the figure)
    and a slow one which drives the liquid motion inside the pores.}
    \label{psi_pres}
\end{figure}
Figure~\ref{psi_pres} shows the temporal evolution of the pore pressure change
calculated with the parameters of the fit at 20 $^\circ C$. It is interesting to note
that the liquid pressure change is never equal to zero (apart when it changes sign)
during the signal evolution. This behavior is different from what appears in a bulk
liquid sample where the pressure equilibrates after the vanishing of the acoustic
oscillations. In the case of confined water, the liquid expansion due to the heating is
partially prevented by the stiffness of the matrix which exerts a pressure on the liquid.
At first, the liquid will change its density via the outflow from the heated pores to
nullify this pressure. At later times, owing to the vanishing of the thermal grating for
thermal diffusion, the matrix will exerts a negative pressure on the liquid which
backflows to the pores to equilibrate again the pressure. Thus, density changes related
to the liquid flow take place on time scales much longer and the liquid pressure is
different from zero at all the times. Finally, the liquid pressure shows two relaxations,
a fast one which drives the acoustic oscillations (the fast Biot wave) and a slow one
which drives the liquid motion inside the pores (the slow Biot diffusive wave).

\section{Fitting Results}\label{sec_Discuss}
In this last section, we shall show and discuss the main findings derived by the
fit-data analysis. We will start with the results of acoustic wave propagation after
that we will analyze the features connected with liquid viscous flow and finally with
the thermal diffusion.

As already said in previous section, many of the parameters appearing in the
equations of P-D model have been kept fixed to the literature values. These values
are, clearly, measured for the bulk materials. For the case of water the assumption
to consider the thermodynamic and dynamic parameters of confined water equal to the
bulk ones is not so obvious. The strong hydrogen bonds of water molecules with the
silanol groups of the inner pore surfaces of Vycor could affect parameters like the
density, specific heat, viscosity, thermal expansivity, etc. For example, it has been
found by molecular simulation~\cite{Gallo2000} that the density of water confined in
Vycor is around $11\%$ lower than the bulk value. Yet, the specific heat has been
measured $1-2\%$ higher than in the bulk~\cite{Tombari2005}. Contrary, measurements
of viscosity of water confined in nanometer films proved that this parameter remains
close to the bulk value~\cite{Raviv2001,Raviv2004}. Nevertheless, the data about the
thermodynamic parameters of water in confined state are few and, in particular, no
data about their temperature evolution exist. For this reason, we have chosen to fix
the values of these parameters to the bulk ones as it has been done so far in all the
experimental study on confined water.

\begin{figure}[t]
\centering
    \includegraphics[scale=0.5]{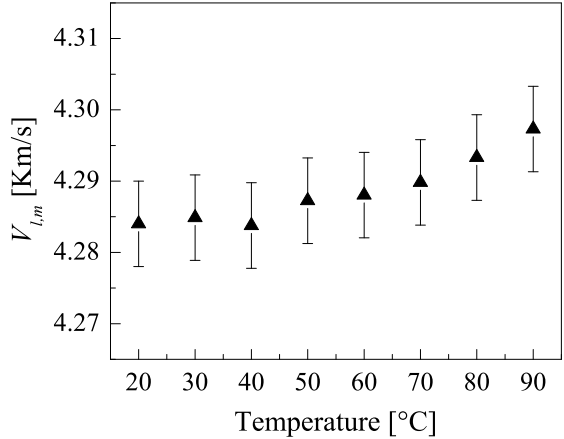}
    \caption{Longitudinal sound velocity of vycor as a function of temperature.}
    \label{Vml}
\end{figure}

In Fig.~\ref{Vml} we report the temperature behavior of the longitudinal sound
velocity of Vycor $V_{l,m}$. We were forced to leave this parameter as free fitting
parameter in order to have a good fit of acoustic oscillations at all the temperatures.
These values together with the transverse velocity values, $V_{t,m}=0.62 V_{l,m}$, and
the values of matrix density, give the matrix stiffness values which differs from the
those measured by static experiments~\cite{Vadakan2000,Vadakan2000b,Corning2008}. But our
longitudinal sound velocities are in reasonable agreement with the values measured by
Brillouin scattering experiments by Levelut and Pelous~\cite{Levelut2007}.
\begin{figure}[t]
\centering
    \includegraphics[scale=0.5]{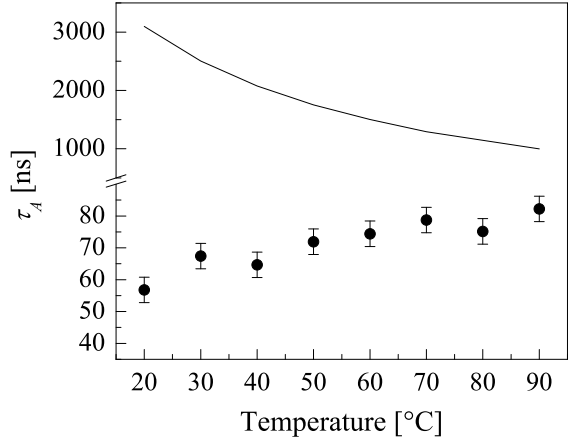}
    \caption{Measured acoustic attenuation times (circles)
    compared with the values predicted by the model of Pecker and Deresiewicz (continuous line).}
    \label{tauA}
\end{figure}

In Fig.~\ref{tauA} we show the measured acoustic attenuation times compared to the
prediction of the P-D model. We see that the times predicted by the theoretical model are
much longer than the experimental ones and moreover their temperature behavior show an
opposite trend. As stated before, P-D model does not introduce substantial modifications
for the wave propagation and retains the same limitations of the Biot model. As just
reported in~\cite{Taschin2008}, the only damping mechanism considered in the theory (the
energy dissipation due to the viscous friction between the matrix and liquid) is not
surely the main damping source in Vycor filled with liquids at our TG frequencies. Other
damping effects should be included, beginning with the dissipation phenomena intrinsic of
the Vycor and water. Considering the very low sound attenuation in bulk water, the
damping could be mainly due to the sound damping in Vycor. Indeed, the underestimations
of the damping mechanism has been already reported in the
literature~\cite{Dvorkin1993,Kibblewhite1989,Williams2002}.

\begin{figure}[t]
\centering
    \includegraphics[scale=0.5]{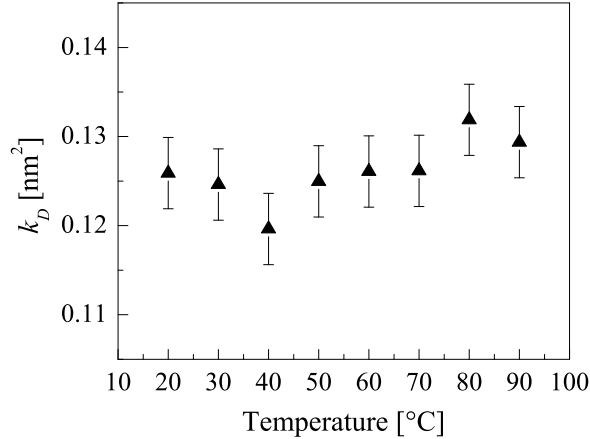}
    \caption{Permeability of water in Vycor extracted from the fits.}
    \label{kD}
\end{figure}

In Fig.~\ref{kD} we show the measured permeability $k_D$ as a function of
temperature. $k_D$ does not show, within the experimental errors, a temperature
dependence and has a mean value of $\sim0.126~nm^2$. The permeability is related to the
viscous Darcy coefficient $b=\phi^2\mu_l/k_D$ and to the hydraulic diffusivity
coefficient~\cite{Smeulders2005}: $D_H=(PR-Q^2)/[b(P+R+2Q)]$, being $P$, $Q$ and $R$ the
generalized elastic moduli defined in the appendix. In fact, the P-D model is an
hydrodynamical theory that supposes that the flow of liquid through the pores is a Darcy
flow (i.e. that the flow is laminar and the fluid velocity at pore wall is zero). This is
assumed in the definition of the dissipative force $b(\dot{e}-\dot{\mathcal{E}})$ and, in
particular, in the expression of $b$. In this model, the permeability of the porous
material, $k_D$, is a quantity depending only on the geometrical characteristics of the
porous material, porosity, mean pore size and tortuosity. Indeed the $k_D$ values
extracted by the fit does not show any appreciable temperature dependence.

The value of the measured permeability can be compared with that of
reference~\cite{Vadakan2000,Vadakan2000b} ($k_D=0.065~nm^2$) and with that calculated
following the expression of the Poiseuille permeability for a porous
medium~\cite{Guo1994}: $k_D=(1/8)r^2\phi/\tau$ where we recall $r$ and $\tau$ are the
mean pore radius and the tortuosity of the medium. By inserting the value of $31\%$ for
the porosity and of $2~nm$ for the mean pore radius, supplied by the Corning company, and
a tortuosity value of 2-4 \cite{Lin1992,Vadakan2000,Vadakan2000b}, we find a permeability
of $0.04-0.08~nm^2$. Thus, our $k_D$ fitting values do not agree with the previous
reported measurements. This could be ascribed to a mean pore size larger than what stated
by the company. Frequently in literature, measured values of porosity and mean pore size
have been found different from the usual Corning values. For example, in ref.
\cite{Vadakan2000,Vadakan2000b}, it is reported a mean pore radius of $2.68~nm$ in
\cite{Li2000} of $2.3~nm$. Our results would be compatible with a mean pore radius around
$3~nm$.

\begin{figure}[t]
\centering
    \includegraphics[scale=0.5]{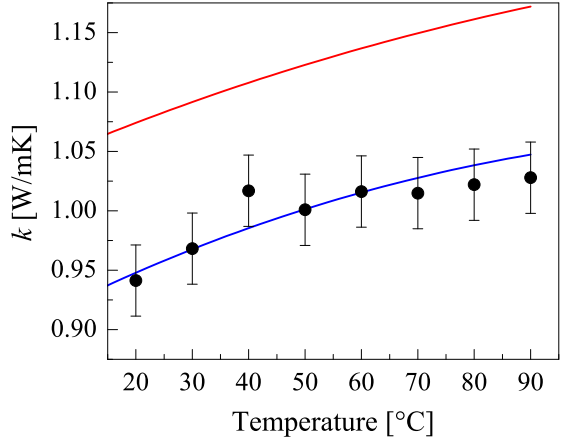}
    \caption{Effective thermal conductivity as a function of temperature
    compared with the maximum theoretical values (red line) and the minimum ones (blue
    line).}
    \label{lamdam}
\end{figure}

In Fig.~\ref{lamdam} we report the temperature behavior of the measured effective
thermal conductivity, $k$. We report also the values of the thermal conductivities
calculated according to the weighted on porosity arithmetic mean (red line),
$k=(1-\phi)k_s+\phi k_l$, and harmonic mean (blue line), $1/k=(1-\phi)/k_s+\phi/k_l$,
being $k_s$ the solid thermal conductivity equal to $1.29$ $W m^{-1}K^{-1}$ @ $20~^\circ
C$ and $k_l$ water thermal conductivity equal to $0.60$ $W m^{-1}K^{-1}$ @ $20~^\circ
C$~\cite{Handbook70}. These expressions of the effective conductivity refers to a medium
composed by parallel layers of solid and water where the heat transfer goes along the
layer direction (parallel conduction) or goes perpendicular to the layer direction
(series conduction)~\cite{AdHT2006}. The value of these two values can be considered as
the two limiting values of the $k$ parameter, see also appendix. The values obtained by
our fitting is in substantial agreement with the harmonic mean.

\begin{figure}[t]
\centering
    \includegraphics[scale=0.5]{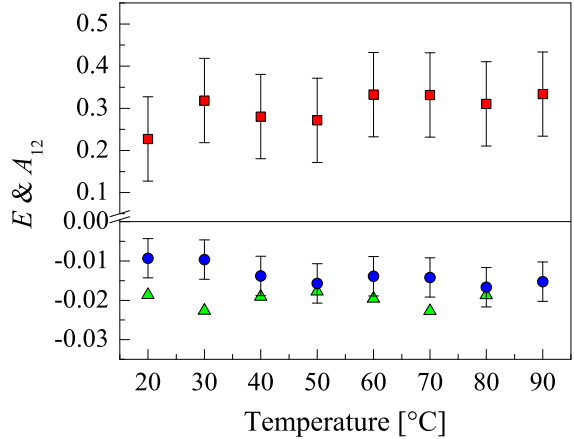}
    \caption{Temperature behavior of the parameters $E$ (blue circles) and $A_{12}$ (red squares).
    Green triangles refer to the $E$ values measured in the bulk water~\cite{Taschin2006}.}
    \label{E12E}
\end{figure}

Finally, we want to show the results obtained for the amplitude parameters $E$ and
$A_{12}$. These are shown in Fig. \ref{E12E}. Both the parameters do not show, within
the experimental error bars, a sensible temperature dependence. In the same figure, the
$E$ values are compared with the ones obtained for the bulk water~\cite{Taschin2006}.
Unfortunately, we have not enough sensibility to estimate if and how the confinement
affects the photothermal effect of water. Anyway, we want to underline that both the
matrix and temperature contributions are absolutely necessary to describe correctly the
data.

Considering the complexity of the fitting model, the agreement of the P-D parameters with
the literature data is reasonable even if it is not complete. Correction on some fixed
parameter could lead probably to a better agreement. In particular, possible changes of
the water parameters due to the confinement should be taken in to account. Nevertheless,
the fact that the model is able to reproduce the data at all the temperatures with a
substantially small number of free parameters catching the main temperature dependencies
and the order of magnitude of the coefficient values, proves the overall validity of the
hydrodynamic laws in describing an nano-heterogenous system.

\section{Conclusions}

In summary, heterodyne detected transient grating experiment has been applied to study
the relaxation dynamic of water confined in Vycor 7930. We acquired HD-TG data in the
temperature range $20-90$ $^\circ C$ at the q-vector of 1 $\mu m^{-1}$ and for the range
of $q=0.63-2.5$ $\mu m^{-1}$ at 30 $^\circ C$. The HD-TG experiment enables to
investigate in a single data the damped acoustic oscillations taking place at short
times, the mass/density dynamics and the thermal diffusion. The HD-TG data has been
analyzed using the hydrodynamic model of thermo-poroelasticity on porous media of Pecker
and Deresiewicz~\cite{Pecker1973}. This hydrodynamic model gives a very interesting
insight of the experimental results enabling a consistent characterization of the
transport processes as they appear in the experimental signal. We need to redefined few
parameters of the model in order to reproduce correctly our experimental data. In
particular the acoustic damping rate that the P-D model, as well as the Biot theory,
underestimate drastically and the effective thermal conductivity that can not described
as a simple weighted mean. The simulation of HD-TG data according to the P-D model shows
that acoustic propagation, taking place on the fast time scale, is due to an in-phase
motion of the solid and liquid components of the system. The mass/density transport is
due to the flow of the confined liquid that results to be coupled to the heat diffusion.
A peculiar flow and back-flow of water inside the nano-pores forced by thermal processes
has been reported. During such transport processes the solid and the liquid are moving
with opposite phases. We fitted our data using the P-D model, fixing as much as possible
the water-Vycor parameters to the known literature values. The values of the free
parameters have been extracted from our data using a best fit procedure. The P-D model
enables a valid fit of our data for the whole investigated time windows, even if there is
not a complete agreement of some fitting values with the values reported in the
literature. In our opinion this disagreement is not to be ascribed to a fail of the
hydrodynamic laws but to an oversimplified definition of the solid and liquid basic
parameters.

We would like to stress that HD-TG experiments turns to be able to measure the
complex dynamic processes, taking place in a nano-heterogenous system, over a very
broad time windows and a relatively simple hydrodynamic model seems appropriate to
describe all the transport and dynamic phenomena measured. Our study gives a further
experimental confirmation of the fact that water flow in Vycor is well described by
hydrodynamic laws in spite of the nanometric dimension of pores.

\section*{Acknowledgments}
The research has been performed at LENS of University of Firenze. We thank D. L. Johnson
for the helpful suggestions and discussions. The research has been supported by the EC
grant N. RII3-CT-2003-506350, by CRS-INFM-Soft Matter (CNR) and MIUR-COFIN-2005 grant N.
2005023141-003.

\appendix

\section{The Pecker and Deresiewicz hydrodynamic model}

According to the Pecker and Deresiewicz model the linearized equations describing the
time evolution for the dilatations $e$ and $\mathcal{E}$ and the temperature variations
$\delta T_m$ and $\delta T_l$ are~\cite{Pecker1973}:

\begin{eqnarray}\label{Pecker1}
        \rho_{11}\ddot{e}+\rho_{12}\ddot{\mathcal{E}}+b(\dot{e}-\dot{\mathcal{E}})
        -P\nabla^2e - Q\nabla^2\mathcal{E}
        + R_{11}\nabla^2\delta T_m + R_{12}\nabla^2\delta T_l=0 \label{Pecker11}
\end{eqnarray}
 \vspace{-20pt}
\begin{eqnarray}
        \rho_{12}\ddot{e}+\rho_{22}\ddot{\mathcal{E}}-b(\dot{e}-\dot{\mathcal{E}})
        - Q\nabla^2e -R\nabla^2\mathcal{E} 
        + R_{21}\nabla^2\delta T_m+R_{22}\nabla^2\delta T_l=0 \label{Pecker12}
\end{eqnarray}
 \vspace{-20pt}
\begin{eqnarray}
        F_{11}\dot{\delta T_m}+F_{12}\dot{\delta T_l} + K(\delta T_m-\delta T_l)
        -(1-\phi)k_s\nabla^2 \delta T_m + R_{11}T_0\dot{e}+R_{21}T_0\dot{\mathcal{E}} =0 \label{Pecker13}
\end{eqnarray}
 \vspace{-20pt}
\begin{eqnarray}
        F_{21}\dot{\delta T_m}+F_{22}\dot{\delta T_l}- K(\delta T_m-\delta T_l)
        - \phi k_l \nabla^2 \delta T_l + R_{12}T_0\dot{e}+R_{22}T_0\dot{\mathcal{E}}=0
        \label{Pecker14}
\end{eqnarray}

in which

\begin{itemize}

\item[-]the densities $\rho_{ij}$ are related to the solid and liquid densities, $\rho_s$
and $\rho_l$, by $\rho_{11}=(1-\phi)\rho_s-\rho_{12}$, $\rho_{22}=\phi\rho_l-\rho_{12}$
and $\rho_{12}=(1-\tau)\phi\rho_l$ where $\tau$ is the tortuosity of the
matrix~\cite{Biot1956a,Biot1956b}.

\item[-]$b=\phi^2\mu_l/k_D$, with $\mu_l$ the water dynamic viscosity and $k_D$ the
Vycor permeability. The $b(\dot{e}-\dot{\mathcal{E}})$ is a viscous friction term
arising from the relative motion between the two materials. This term is, at the same
time, the damping source for the acoustic waves and the hindrance to the liquid slow
flow through pores.

The  $P$, $Q$ and $R$ coefficients are the generalized isothermal elastic moduli defined
in the Biot theory~\cite{Biot1956a,Biot1956b}, they can be related to the isothermal bulk
modulus of liquid $K_l$, the bulk modulus of solid $K_s$, the bulk modulus of matrix
$K_m$ and to $N$ which is the isothermal shear modulus of matrix. In particular we
have~\cite{Johnson1982,Carcione2001}
\begin{eqnarray}
    P=\frac{(1-\phi)\left[1-\phi-\frac{K_m}{K_s}\right]K_s+\phi\frac{K_s}{K_l}K_m}
       {1-\phi-\frac{K_m}{K_s}+\phi \frac{K_s}{K_l}}+\frac{4}{3}N\\
    Q=\frac{\left(1-\phi-\frac{K_m}{K_s}\right)\phi K_s}{1-\phi-\frac{K_m}{K_s}+\phi \frac{K_s}{K_l}}\\
    R=\frac{\phi^2 K_s}{1-\phi-\frac{K_m}{K_s}+\phi \frac{K_s}{K_l}}
\end{eqnarray}
with $\phi$ the matrix porosity.

We recall that $K_l$, $K_s$, $K_m$ and $N$ can be expressed as functions of the density
and the longitudinal and transverse sound velocities
\begin{eqnarray}
    K_l=\rho_lV_{l}^2/\gamma_r \,, \; K_s=\rho_s(V_{l,s}^2-4/3V_{t,s}^2) \,,\nonumber\\
    K_m=\rho_m(V_{l,m}^2-4/3V_{t,m}^2) \,, \; N=\rho_m V_{t,s}^2 \nonumber
\end{eqnarray}

The coefficients $R_{ij}$ are the generalized expansion coefficient defined by
following equations~\cite{Johnson1982,Carcione2001}:
\begin{eqnarray}
    R_{11}=\alpha_s(3P-4N)+\alpha_{ls}Q \nonumber \\
    R_{12}=\alpha_{sl}(3P-4N)+\alpha_{l}Q\nonumber \\
    R_{21}=3\alpha_{s}Q+\alpha_{ls}R \nonumber \\
    R_{22}=3\alpha_{sl}Q+\alpha_{l}R
\end{eqnarray}
where $\alpha_s$ is the linear expansion coefficient of solid and $\alpha_l$ is the
volumetric thermal expansion coefficient liquid, $\alpha_{sl}$ and $\alpha_{ls}$ are
the thermo-elastic coupling coefficients.

The $F_{ij}$ coefficients the are generalized specific heats at constant volume and
they are defined by the following expressions~\cite{Johnson1982,Carcione2001}:
\begin{eqnarray}
    F_{11}=(1-\phi)\rho_s C_{ps}-T_0(3\alpha_s R_{11}+\alpha_{ls} R_{21}) \nonumber \\
    F_{12}=-T_0(3\alpha_s R_{12}+\alpha_{ls} R_{22})\nonumber \\
    F_{21}=-T_0(3\alpha_{sl} R_{11}+\alpha_{l} R_{21}) \nonumber \\
    F_{22}=\phi\rho_l C_{pl}-T_0(3\alpha_{sl} R_{12}+\alpha_{l} R_{22})
\end{eqnarray}
where $C_{ps}$ and $C_{pl}$ are the isobaric specific heats of solid and liquid and
$T_0$ the equilibrium temperature of the whole system.

\item[-]$K$ is the coefficient of interphase heat transfer.

\item[-]$k_s$ and $k_l$ are the thermal conductivities of solid and liquid.

\end{itemize}

Now some approximations can be made to simplify the equations. As already discussed in
section \ref{sec_Exp_Res}, we can assume the local thermal equilibrium between the
two phases. This implies the assumption of a $K$ coefficient very high in the equations
and thus $T_m=T_l=T$. An other hypothesis is that the thermo-elastic coupling
coefficients, $\alpha_{sl}$ and $\alpha_{ls}$ can be considered negligible. This is a
plausible approximation considering the very low thermal expansion coefficient of Vycor
and its stiffness. In fact, $\alpha_{sl}$ is the strain in the matrix due to an unit
change of temperature in the liquid and $\alpha_{ls}$ is the dilatation of the liquid due
to an unit change of temperature in the matrix. We expect that the former is negligible
due to the stiffness of the matrix ($K_m=13.4~GPa>K_l=2.2~GPa$ @20 $^\circ C$) and the
latter is negligible due to the low expansivity of the matrix ($3\alpha_s=2.25\times
10^{-6}~K^{-1}<<\alpha_l=2.07\times10^{-4}~K^{-1}$ @20 $^\circ C$). After these
approximations and summing together Eq.s (\ref{Pecker13}) and (\ref{Pecker14}) in order
to obtain a single equation for the temperature $T$, the system of
equations~(\ref{Pecker1}) becomes

    \begin{eqnarray}
        \rho_{11}\ddot{e}+\rho_{12}\ddot{\mathcal{E}}
        +b(\dot{e}-\dot{\mathcal{E}})-P\nabla^2e-Q\nabla^2\mathcal{E}
        +R_1\nabla^2\delta T=0
    \end{eqnarray}
 \vspace{-20pt}
    \begin{eqnarray}
        \rho_{12}\ddot{e}+\rho_{22}\ddot{\mathcal{E}}
        -b(\dot{e}-\dot{\mathcal{E}})-Q\nabla^2e-R\nabla^2\mathcal{E}
        +R_2\nabla^2\delta T=0
    \end{eqnarray}
 \vspace{-20pt}
    \begin{eqnarray}
        F\dot{\delta T}+R_1T_0\dot{e}+R_2T_0\dot{\mathcal{E}}
        -k\nabla^2 \delta T=0 \label{Pecker2}
    \end{eqnarray}
where
\begin{eqnarray}
    R_1=\alpha_s(3P-4N)+\alpha_lQ \nonumber\\
    R_2=3\alpha_sQ+\alpha_lR. \nonumber\\
    F=\sum F_{ij}=(1-\phi)\rho_s C_{ps}+\phi\rho_lC_{pl}-T_0(3\alpha_sR_1+\alpha_lR_2)\nonumber\\
    k=(1-\phi)k_s+\phi k_l \label{Pecker3}
\end{eqnarray}
P-D model, in its original form, predicts an effective thermal conductivity, $k$,
defined by an arithmetic means of values of the two constituent phases, weighted on
porosity: $k=(1-\phi)k_s+\phi k_l$. This kind of mean is used also in the definition
of the effective volumetric heat capacity appearing in the expression of $F$
coefficient. Since the effective specific heat is independent from the medium
morphology, this definition appears appropriate and correct. Whereas, it is well
known that the thermal conductivity is generally dependent by the medium
morphology~\cite{AdHT2006}, in the present sample by the geometric characteristics of
the porous material. The expression reported in eq.~\ref{Pecker3} refers to a medium
composed by parallel layers of solid and water where the heat transfer goes along the
layer direction (parallel conduction). Clearly, this picture represents a very
particular case which, anyway, defines a maximum limit for the effective
conductivity. The minimum limit is, instead, obtained with the harmonic mean,
$1/k=(1-\phi)/k_s+\phi/k_l$, which describes a composite medium of parallel layers
where the heat transfer goes perpendicular to the layer direction (series
conduction). In a generic porous material, where the tubules are interconnected and
randomly oriented in all directions, the effective thermal conductivity should assume
values within these two limits~\cite{Woodside1961a,Woodside1961b,AdHT2006}.

Since the TG signal is directly related to a $q$-component of spatial Fourier transform
of the dielectric constant change~\cite{CapBook_TG2008}, the next step is writing in the
$q$-space the system of equations~(\ref{Pecker2}):
\begin{eqnarray}
    \rho_{11}\ddot{e}+\rho_{12}\ddot{\mathcal{E}}
    +b(\dot{e}-\dot{\mathcal{E}})+q^2Pe+q^2Q\mathcal{E}-q^2R_1\delta T=0 \nonumber\\
    \rho_{12}\ddot{e}+\rho_{22}\ddot{\mathcal{E}}-b(\dot{e}-\dot{\mathcal{E}})+q^2Qe+q^2R\mathcal{E}-q^2R_2\delta T=0 \nonumber\\
    F\dot{\delta T}+R_1T_0\dot{e}+R_2T_0\dot{\mathcal{E}}+q^2k \delta T=0 \label{Pecker33}
\end{eqnarray}
where, now, $e$, $\mathcal{E}$ and $\delta T$ refer to a $q$-component of the their
spatial fourier transforms.

\end{document}